\documentclass[pra,aps,twocolumn,epsfig,superscriptaddress,showpacs]{revtex4}
\usepackage{bm}
\usepackage{amsfonts}
\usepackage[dvips]{graphicx}
\usepackage{mathrsfs}
\usepackage[intlimits]{amsmath}
\usepackage[colorlinks=false,dvipdfm]{hyperref}

\begin{document}

\author{Dong-Ling Deng}
 \affiliation{Theoretical Physics Division, Chern Institute of
Mathematics, Nankai University, Tianjin 300071, People's Republic of
China}
\author{Zi-Sui Zhou}
\affiliation{College of Mathematics, Nankai University, Tianjin
300071, People's Republic of China}
\author{Jing-Ling Chen}
 \email{chenjl@nankai.edu.cn}
\affiliation{Theoretical Physics Division, Chern Institute of
Mathematics, Nankai University, Tianjin 300071, People's Republic of
China}

\date{\today}

\title{Relevant multi-setting tight Bell inequalities for qubits and qutrits}

\begin{abstract}
In the celebrated paper [J. Phys. A: Math. Gen. 37, 1775 (2004)], D.
Collins and N. Gisin presented for the first time a three setting
Bell inequality (here we call it CG inequality for simplicity) which
is relevant to the Clauser-Horne-Shimony-Holt (CHSH) inequality.
Inspired by their brilliant ideas, we obtained some multi-setting
tight Bell inequalities, which are relevant to the CHSH inequality
and the CG inequality. Moreover, we generalized the method in the
paper [Phys. Rev. A 79, 012115 (2009)] to construct Bell inequality
for qubits to higher dimensional system. Based on the generalized
method, we present, for the first time, a three setting tight Bell
inequality for two qutrits, which is maximally violated by
nonmaximally entangled states and relevant to the Collins-Gisin-
Linden-Massar-Popescu inequality.
\end{abstract}

\pacs{03.65.Ud, 03.67.Mn, 03.65.Ca} \maketitle

\section{Introduction}
Two Bell inequalities (BIs) are said to be relevant if there exist
quantum states that violate one of them but not the other.
Analogously, an inequality is said to be relevant to a given set of
inequalities if there exist quantum states violating it, but not
violating any of the inequalities in the set~\cite{N.Gisin2}. A
long-surviving open question proposed by N. Gisin is
that~\cite{N.Gisin2}: for a given Hilbert space whose dimension is
limited, is there a finite set of inequalities such that no other
inequality is relevant with respect to that set? The significance
considering this problems are two-folded: On  the one hand, it helps
to reveal the recondite relationship between various multi-setting
BIs. On the other hand, since BIs are at the heart of the study  of
non-locality and many rather nonintuitive quantum phenomena such as
quantum key secret sharing~\cite{V Scarani}, quantum communication
complexity problems~\cite{C Brukner}, etc., can be measured with BIs
of some form, it may contribute a lot to the field of quantum
information and computation.

Historically, Bell inequality, which may rule out all local
hidden-variables theories, was first proposed by J. S. Bell in
$1964$~\cite{J.S.Bell} when studying the Einstein-Podolsky-Rosen
(EPR) paradox~\cite{A. Einstein-EPR}. After his pioneering work,
there arose many generalization of the original Bell inequality.
Among all of these generalizations, the Clauser-Horne-Shimony-Holt
(CHSH) inequality~\cite{J.Clauser}, which is a two-setting
inequality for two-qubit system, may be the most famous one. The
CHSH inequality has many merits: (i) it is tight,  i.e., it defines
one of the facets of the convex polytope~\cite{A.Peres-1998} of
local-realistic (LR) models; (ii) it violate all the pure two-qubit
entangled states; (iii) it is maximally violated by maximally
entangled states. Another significant step to generalize the Bell
inequality for two $d$-dimensional (qudit) system is made by D.
Collins \textit{et al}, who constructed a CHSH type inequality for
arbitrary $d$-dimensional (qudit) systems  in $2002$, now known as
the Collins-Gisin-Linden-Masser-Popescu (CGLMP)
inequality~\cite{2002-CGLMP-Inequality}. This inequality was shown
to be tight~\cite{L.Masanes} but it doesn't preserve the merit (iii)
for CHSH inequality, i.e., it's maximal violation occurs at the
nonmaximally entangled states~\cite{2006Chen-MaxV}. It is worthwhile
to note that, more recently, a new tight inequality preserve this
merit was proposed by S. W. Lee and D. Jaksch~\cite{2008Lee-Optimal
BI}.

All the inequalities mentioned above belong to the two-setting Bell
inequalities, i.e., they are based on the standard Bell experiment,
in which each local observer is given a choice between two
dichotomic observables. However, one may extend the number of
measurement settings. Actually, multi-setting Bell inequalities may
have many advantages in many protocols in quantum information
theory. Then, to constructing new multi-setting BIs is an important
and pressing work.

%
%
%

Let's go back to the open question mentioned in the first paragraph.
The first step concerning this problem was made by D. Collins and N.
Gisin. In $2004$, they presented for the first time a three setting
tight Bell inequality (here we call it CG inequality for simplicity)
which is relevant to the CHSH
inequality~\cite{D.Collins-N.Gisin-Rel}. In this paper, we
generalized the method introduced in Ref.~\cite{2008Chen-HolderToBI}
for constructing Bell inequality to higher dimensional systems.
Using this method, we obtain  some multi-setting tight Bell
inequalities, which are relevant to the CHSH inequality and the CG
inequality. Moreover, we present, for the first time, a three
setting tight Bell inequality for two qutrits, which is maximally
violated by nonmaximally entangled states and relevant to the CGLMP
inequality.

The article is organized as follows. In Sec. II, we first review the
method for constructing Bell inequality for qubits introduced in
Ref.~\cite{2008Chen-HolderToBI} and then generalize this method to
higher dimensional system. In Sec. III, we presented various
multi-setting tight Bell inequalities for two-qubit systems. In Sec.
IV, a three-setting tight Bell inequality, which is maximally
violated by nonmaximally entangled state and relevant to the CGLMP
inequality, is introduced. Finally in Sec. V, we concluded the
article with some final remarks.

\section{An effective method for constructing Bell inequalitly}
Finding all the Bell inequalities for a given number of measurement
settings and outcomes is a very difficult
problem~\cite{1989Pitowski}. There exist various method to construct
Bell inequalities~\cite{1990Mermin}. In this section, we first
review the method for constructing Bell inequality for qubits
introduced in Ref.~\cite{2008Chen-HolderToBI} and then generalize it
to higher dimensional systems case. In
Ref.~\cite{2008Chen-HolderToBI}, the author develop a systematic
approach to establish Bell inequalities for qubits based on the
Cauchy-Schwarz inequality. Let's consider the following Bell-type
scenario: $N$ spatially separated observers measures independently
among $M$ observables with two possible outcomes $-1$ and $1$,
determined by some local parameters denoted by $\lambda$. We denote
by $X_j(\mathbf{n}_{k_j},\lambda)$, or $X_{j,k_j}$ for simplicity,
the observables on the $j$-th party.
Then the correlation function, in the case of a local realistic
theory, is defined as
$Q(\mathbf{n}_{k_1},\mathbf{n}_{k_2},\cdots,\mathbf{n}_{k_j})=\int_\Gamma
\Pi_{j=1}^NX_j(\mathbf{n}_{k_j},\lambda) \rho(\lambda) d\lambda$,
where $j=1,2,\cdots,N$ and $k_j=1,2,\cdots,M$. For convenience, we
denote the correlation function
$Q(\mathbf{n}_{k_1},\mathbf{n}_{k_2},\cdots,\mathbf{n}_{k_j})$ as
$Q_{k_1k_2\cdots k_j}$. Define the Bell function as:
\begin{eqnarray}\label{general-BF}
\mathcal {B}(\lambda)=\sum_{\chi}C_{\chi}\prod_{j=1}^{N}X_{j,k_j}.
\end{eqnarray}
Note that the symbol $\chi$ associated with
$\prod_{j=1}^{N}X_{j,k_j}$ stands for $N$ pairs of indices (one pair
for each observer). Obviously, there are $N_\chi=(1+M)^N$ distinct
values of $\chi$. The constant numbers $C_{\chi}$s are coefficients
of $\prod_{j=1}^{N}X_{j,k_j}$. In a local realistic theory, for each
set of values of $X_j$s, the Bell function corresponds to a number,
which is called a ``\emph{root}" of the Bell function ${\cal
B}(\lambda)$. We now rewrite the theorem $1$ and $2$ in
Ref.~\cite{2008Chen-HolderToBI} as a single theorem without proof.
Anyone who want to know the details please see
Ref.~\cite{2008Chen-HolderToBI}.

\emph{Theorem 1.}  Let $S_2=\{\mathcal {B}(\lambda) \;|\; \mathcal
{B}^2(\lambda)=1\}$, i.e., $\mathcal {B}(\lambda)$ must have two
distinct ``\emph{roots}" $\Lambda_1=-1$ and $\Lambda_2=1$, then for
$\forall \; \mathcal {B} (\lambda)\in S_2$, one has the Bell
inequality: $|\langle \mathcal {B} (\lambda)\rangle_{LHV}|\leq 1$.
In general, if  $ S_n=\{\mathcal {B}(\lambda) \;|\;
\prod_{j=0}^{j=n-1} (\mathcal {B}-\Lambda_j)=0,\;\mathcal
{B}(\lambda)\notin \bigcup_{k=2}^{k=n-1}S_k, \; n\in {\rm integers},
\; n \ge 3\}$, which means that $n$ ``\emph{roots}" of $\mathcal
{B}(\lambda)$ uniformly distribute between $-1$ and $1$ with
$\Lambda_j=-1+2j/(n-1)$, for $\forall \; \mathcal {B} (\lambda)\in
S_n$, one has the Bell inequality: $|\langle \mathcal {B}
(\lambda)\rangle_{LHV}|\leq 1$.

Based on this theorem, we only need to determine the coefficients
$C_{\chi}$ in Eq.~(\ref{general-BF}) so that the Bell function
$\mathcal {B}(\lambda)$ has only specified distinct ``\emph{roots}".
This can be done by solve a series of equations of $C_{\chi}$s. To
generalize this method to $N$-qudit system, where the observable
$X_{j,k_j}$ on the $j$-th party has $d$ possible outcomes $0$, $1$,
$\cdots$, $d-1$, we should use the joint probability instead of the
correlation function for the snake of simplicity and conveniency.
Denote by $P(X_{1,1_1}=x_{1,1_1}, \cdots,X_{N,M_N}=x_{N,M_N})$ the
joint probability that the $j$th party measures $X_{j,k_j}$ and
obtains the outcome $x_{j,k_j}$, then the general Bell function for
$N$-qudit system can be defined as:
\begin{eqnarray}\label{general-BF-Prob}
\mathcal {B}(\lambda)=\sum_{\chi}C_{\chi}P(X_{1,1_1}=x_{1,1_1},
\cdots,X_{N,M_N}=x_{N,M_N}).
\end{eqnarray}
After this modification of the Bell function, one can also use the
theorem $1$ to construct Bell inequality for $N$-qudit systems and
the procedures are the same as that of qubits case.

This method is efficient to construct various multi-setting Bell
inequality. However, like many other method to construct Bell
inequalities, it has its own disadvantages: (i) the equations of
$C_{\chi}$s are Pan-set equations, i.e., the number of independent
equations is less than the number of variables $C_{\chi}$s. Thus it
is difficult to find all the solves of this equations; (ii) Some
inequalities obtained by this method are trivial, i.e., they cannot
be violated in quantum mechanics; (iii) This method gives no
information about the tightness of the obtained Bell inequalities.
Plentiful numerical results provide an empirical conclusion: the
less the number of distinct ``\emph{roots}" of the Bell function,
the more likely that the corresponding Bell inequality is tight.
(iv) The number of unknown coefficients $C_{\chi}$ grows
exponentially with the dimension of the system. Thus, when the
system get large, it is very difficult to solve the equations of
$C_{\chi}$s. Actually, in order to simplify the calculation, we
always choose only these Bell functions which possess some nature
symmetry but not the general one as defined in equations
~(\ref{general-BF}) or (\ref{general-BF-Prob}).


\section{multi-setting tight Bell inequality for two qubits}
In this section, we shall focus on Bell inequality for two-particle
systems. The Bell-type scenario involves only two observers and each
of them measures $M$ different local observables of two outcomes
$\pm1$. For simplicity and convenience, we denote $X_{1,1_k}$,
$X_{2,2_k}$ as $A_k$ and $B_k$ ($k=1,\cdots,M$) respectively.
%
%
The correlation function $Q(A_iB_j)$,  in the case of a local
realistic theory, is then the average values of the products
$A_iB_j$ over many runs of the experiment. We also denote
$Q(A_iB_j)$, $Q(A_i)$ and $Q(B_j)$ as $Q_{ij}$, $Q_{i0}$ and
$Q_{0j}$, respectively.  Then the famous CHSH inequality:

\begin{eqnarray}\label{CHSH}
I_{CHSH}=Q_{11}+Q_{12}+ Q_{21}-Q_{22}\leq2.
\end{eqnarray}
holds in any local realistic theory. The CHSH inequality is almost
always the most efficient one to prove a quantum state to be
nonlocal. The first Bell inequality relevant to the CHSH inequality
was proposed by D. Collins and N. Gisin in $2004$
\cite{D.Collins-N.Gisin-Rel}. In the form of joint probability,
their inequality reads:
\begin{widetext}
\begin{eqnarray}
I_{CG}&=&P(a_1=0,b_1=0)+P(a_1=0,b_2=0) +P(a_1=0,b_3=0)
+P(a_2=0,b_1=0) +P(a_3=0,b_1=0)
\nonumber\\&-&P(b_2=0)+P(a_2=0,b_2=0)-P(a_2=0,b_3=0)-P(a_3=0,b_2=0)-P(a_1=0)
-2P(b_1=0)\leq0
\end{eqnarray}
\end{widetext}
It is easy to check that the CG inequality is tight and its
corresponding Bell function has four distinct \emph{``roots"}. For
state:
\begin{eqnarray}\label{cg-state}
\rho_{cg}=0.85P_{|\psi\rangle}+0.15P_{|01\rangle}.
\end{eqnarray}
where $|\psi\rangle=\frac{1}{\sqrt{5}}(2|00\rangle+|11\rangle)$, one
can check that this state does not violate the CHSH inequality.
However it does violate the CG inequality and the violation is
$0.0129$.

Inspired by their brilliant ideas, we find various Bell
inequalities, which are relevant to the CHSH inequality and the CG
inequality. Our approach to the new Bell inequalities are based on
the method introduced in Sec. II. To illuminate how this method
works, here we show the details for the constructing of a
three-setting BI. Firstly, we define a three-setting Bell function
as:
\begin{eqnarray}
\mathcal{B}(\lambda)&=&c_0+c_1(A_1+B_1)+c_2(A_2+B_2)+c_3(A_3+B_3)\nonumber\\
&+&c_4A_1B_1+c_5(A_1B_2+A_2B_1)+c_6(A_1B_3+A_3B_1)\nonumber\\
&+&c_7A_2B_2+c_8(A_2B_3+A_3B_2)+c_9A_3B_3.
\end{eqnarray}
Here, we assume that the Bell function is symmetric under the
permutations of $A_i$ and $B_i$. Then, in order to find a inequality
whose corresponding Bell function may have four distinct
\emph{``roots"}, we should calculate
$\mathscr{L}(\lambda)=(\mathcal{B}^2(\lambda)-1)(\mathcal{B}^2(\lambda)-1/9)$
and simplify the result using the equations $A_i^2=1$ and $B_i^2=1$.
Let all the coefficients in the simplified expression of
$\mathscr{L}(\lambda)$ equal to zero and solve the equations.
Finally, one obtain a series of solutions for the coefficients $c_j$
($j=0,\cdots,9$). Each solution corresponds to a Bell inequality.
However, some of these inequalities are trivial or equivalent, thus
we have to rule out them by calculate the quantum violation and
check the tightness of each inequality. After some direct
calculation, our new three-setting BI is as follows:
\begin{eqnarray}\label{3settingRBI}
-8&\leq&I_3^4= Q_{21}+Q_{12}+Q_{31}+Q_{13}
+Q_{32}+Q_{23}\nonumber\\
&-&Q_{11}-Q_{22}+Q_{10}+Q_{01}-Q_{20}-Q_{02}\leq4.
\end{eqnarray}
In quantum mechanics, the observables could be spin projections onto
unit vectors $A_i=\mathbf{n}_{a_i}\cdot\vec{\sigma}$,
$B_j=\mathbf{n}_{a_j}\cdot\vec{\sigma}$, and for a two-qubit state
$\rho$, we have:
$Q(A_iB_j)=\texttt{Tr}[(\mathbf{n}_{a_i}\cdot\vec{\sigma}\otimes\mathbf{n}_{b_j}\cdot\vec{\sigma})\rho]$.
$Q(A_0B_j)=\texttt{Tr}[\mathbb{I}\otimes\mathbf{n}_{b_j}\cdot\vec{\sigma})\rho]$
and $Q(A_iB_0)=\texttt{Tr}[(\mathbf{n}_{a_i}\cdot\vec{\sigma}\otimes
\mathbb{I})\rho]$. Here, $\mathbb{I}$ is the $2\times 2$ identity
matrix and $\vec{\sigma}$ is the vector of Pauli matrices.
Inequality~(\ref{3settingRBI}) is tight and also have four distinct
``\emph{roots}". Its maximal quantum violation occurs at the
maximally entangled state:
$|\psi\rangle_2^{max}=\frac{1}{\sqrt{2}}(|00\rangle+|11\rangle)$ and
the the violation is $5$. This inequality is relevant to the CHSH
inequality. One can check that the state $\rho_{cg}$ also violate
this inequality and the violation is $4.0516$. In fact, this
inequality is equivalent to the CG inequality.

We also obtain two four-setting tight Bell inequalities. The first
one, which has four distinct \emph{``roots"}, reads:
\begin{eqnarray}\label{4setting4RBI-1}
-6&\leq&I_{[4]}^4=Q_{11}+Q_{22}+Q_{12}+Q_{21}
+Q_{14}+Q_{41}-Q_{24}\nonumber\\
&-&Q_{42} -2Q_{33}+Q_{31}+Q_{13}+Q_{32}+Q_{23}\leq6.
\end{eqnarray}
The inequality~(\ref{4setting4RBI-1}) is relevant to the CHSH
inequality and the CG inequality. In fact, we have find numerically
many states that do not violate the CHSH (or CG) inequality but
violate the inequality~(\ref{4setting4RBI-1}) and vice versa. Here,
for simplicity, we only present a single one that do not violate the
CG inequality but violate the inequality~(\ref{4setting4RBI-1}):
\begin{widetext}
\begin{eqnarray}
\rho_1=\left(\begin{array}{cccc}
0.046125&-0.057737+0.017786i&-0.000649-0.092414i&0.054845+0.071287i\nonumber\\
-0.057737-0.017786i&0.146863&-0.039254+0.242031i&-0.103099-0.199746i\nonumber\\
-0.000649+0.092414i&-0.039254-0.242031i&0.428573&-0.307414+0.244118i\nonumber\\
0.054845-0.071287i&-0.103099+0.199746i&-0.307414-0.244118i&0.378439\end{array}\right).
\end{eqnarray}
\end{widetext}
One may easily check that $\rho_1$ does not violate the CG
inequality but do violate the inequality~(\ref{4setting4RBI-1}) and
the violation is $6.63804$. The maximal violation of the
inequality~(\ref{4setting4RBI-1}) also occurs at the maximally
entangled state $|\psi\rangle_2^{max}$ and the violation is
$8.1655$. Its threshold visibility is $0.7348$, which is larger than
that of the CHSH inequality, indicating that this inequality is not
as strong as the CHSH inequality. The other new four-setting tight
BI with five distinct \emph{``roots"} reads:
\begin{eqnarray}\label{4setting5RBI-2}
-10&\leq&I_{[4]}^{5}=Q_{10}+Q_{01}-Q_{12}-Q_{21}+Q_{33}-Q_{44}\nonumber\\
&+&Q_{31}+Q_{13}+Q_{32}+Q_{23}-Q_{04}-Q_{40}\nonumber\\
&+&Q_{41}+Q_{14}+Q_{43}+Q_{34}\leq6.
\end{eqnarray}
It is surprising that the maximal quantum violation of this
inequality occurs at the nonmaximally entangled state
$|\psi\rangle_2^{nmax}=0.718824|00\rangle+0.695192|11\rangle$ and
the violation is $7.74134$, which is larger than $7.7395$, the
violation of the maximally entangled state $|\psi\rangle_2^{max}$.
One can also check that this inequality is relevant to the CHSH
inequality, the CG inequality and the
inequality~(\ref{4setting4RBI-1}).

It will be useful for later on to write inequality
(\ref{4setting4RBI-1}) in the following way:
%
\begin{eqnarray}\label{4Set-MF}
I_{[4]}=\left(\begin{array}{c||cccc} &A_1&A_2&A_3&A_4\\ \hline\hline B_1&1&1&1&1\\
B_2&1&1&1&-1\\
B_3&1&1&-2&0\\
B_4&1&-1&0&0\\
\end{array}\right).
\end{eqnarray}
Here, the coefficient in the matrix indicate the coefficients of the
corresponding expectation values.

Using the same method, we also found many six-setting Bell
inequality for two qubits. Here we present only one of them whose
correlation coefficients are regular with respect to the CHSH
inequality and inequality~(\ref{4setting4RBI-1}). In the matrix
form, it reads:
\begin{eqnarray}\label{sixSBI}
I_{[6]}=\left(\begin{array}{c||cccccc} &A_1&A_2&A_3&A_4&A_5&A_6\\ \hline\hline B_1&1&1&1&1&1&1\\
B_2&1&1&1&1&1&-1\\
B_3&1&1&1&1&-2&0\\
B_4&1&1&1&-3&0&0\\
B_5&1&1&-2&0&0&0\\
B_6&1&-1&0&0&0&0\\
\end{array}\right).
\end{eqnarray}
Inequality~(\ref{sixSBI}) is also tight and relevant to the previous
inequalities. Its corresponding Bell function has seven distinct
\emph{``roots"}. From our numerical results, it seems like that
multi-setting tight Bell inequalities with different number of
distinct \emph{``roots"} may be always relevant to each other.
Inspired by inequality~(\ref{CHSH}), (\ref{4Set-MF}) and
(\ref{sixSBI}), it is not difficult to guess the general form of a
set of even setting Bell inequalities:
\begin{widetext}
\begin{eqnarray}\label{2nsettingBI}
I_{[2n]}=\left(\begin{array}{c||ccccccccccc}
&A_1&A_2&A_3&\cdots&A_n&A_{n+1}&A_{n+2}&\cdots&A_{2n-2}&A_{2n-1}&A_{2n}\\
\hline\hline B_1&1&1&1&\cdots&1&1&1&\cdots&1&1&1\\
B_2&1&1&1&\cdots&1&1&1&\cdots&1&1&-1\\
B_3&1&1&1&\cdots&1&1&1&\cdots&1&-2&0\\
\vdots&\vdots&\vdots&\vdots&\vdots&\vdots&\vdots&\vdots&\vdots&\mbox{\scalebox{-1}[1]{$\ddots$}}&\mbox{\scalebox{-1}[1]{$\ddots$}}&\vdots\\
B_n&1&1&1&\cdots&1&1&-(n-1)&\cdots&0&0&0\\
B_{n+1}&1&1&1&\cdots&1&-n&0&\cdots&0&0&0\\
B_{n+2}&1&1&1&\cdots&-(n-1)&0&0&\cdots&0&0&0\\
\vdots&\vdots&\vdots&\vdots&\vdots&\vdots&\vdots&\vdots&\vdots&\vdots&\vdots&\vdots\\
B_{2n-2}&1&1&1&\mbox{\scalebox{-1}[1]{$\ddots$}}&0&0&0&\cdots&0&0&0\\
B_{2n-1}&1&1&-2&\mbox{\scalebox{-1}[1]{$\ddots$}}&0&0&0&\cdots&0&0&0\\
B_{2n}&1&-1&0&\cdots&0&0&0&\cdots&0&0&0\\
\end{array}\right)\leq n(n+1).
\end{eqnarray}
\end{widetext}
Numerically, the inequality (\ref{2nsettingBI}) is tight and its
corresponding Bell function has $(n^2+n+2)/2$ distinct
``\emph{roots}". In fact, we have checked it for less than $22$
settings and believe this is true for arbitrary even settings.
However, to prove it analytically seems difficult and we have to
leave it open. It is worthwhile to note that the inequality
(\ref{2nsettingBI}) was also obtained in Ref.~\cite{N.Gisin2} by
searching numerically all possibilities assuming small integer
coefficient of the matrix. Its quantum violation and other
properties were also discussed there. We constructed this set of
Bell inequalities using a fresh method. One of the interests of the
inequality (\ref{2nsettingBI}) is that some of its reduced
inequalities are also tight. For instance, if we set $A_6$ and $B_6$
equal to $1$, then we get a five-setting Bell inequality reduced
from the inequality~(\ref{sixSBI}):

\begin{eqnarray}\label{five-SettingBI}
I_{[5]}^{red}&=&Q_{11}+Q_{12}+Q_{13}+Q_{14}+Q_{15}+Q_{21}+Q_{22}\nonumber\\
&+&Q_{23}+Q_{24}+Q_{25}+Q_{31}+Q_{32}+Q_{33}+Q_{34}\nonumber\\
&-&2Q_{35}+Q_{41}+Q_{42}+Q_{43}-3Q_{44}+Q_{31}\nonumber\\
&+&Q_{32}+Q_{51}+Q_{52}-2Q_{53}+Q_{10}+Q_{01}\nonumber\\
&-&Q_{20}-Q_{02}\leq12.
\end{eqnarray}
It is easy to check that inequality~(\ref{five-SettingBI}) is also
tight. More generally, if we set $A_{2n}$ and $B_{2n}$ in
inequality~(\ref{2nsettingBI}) equal to $1$, then most of, not all,
the reduced inequality $I_{[2n-1]}^{red}\leq n(n+1)$ is tight. In
fact, we have check to $15$ setting and find that only the
inequality $I_{[3]}^{red}$ is not tight, which is surprising. We
guess that all the inequalities $I_{[2n-1]}^{red}\leq n(n+1)$ are
tight for all $n\geq3$. If this is true, then we can obtain a set of
odd setting tight Bell inequalities. Another interesting question
related is whether we can get a BI reduced form
inequality~(\ref{2nsettingBI}) that may stronger than the CHSH
inequality?

\section{multi-setting tight Bell inequality for two qutrits}
Most of multi-setting Bell inequalities are construct for
qubits~\cite{2006K. Nagata,2007M Wiesniak}. Up to now, there is few
multi-setting Bell inequalities for higher dimensional systems. More
recently, Ji \textit{et al}. introduced a three setting Bell
inequality for two-qutrit systems, which is maximally violated by
maximally entangled state if local measurements are configured to be
mutually unbiased~\cite{2008Ji}. Here based on the generalized
method in Sec. II, we construct two new three-setting Bell
inequalities. One of them is tight and relevant to the CGLMP
inequality. The other one is not tight but it is as strong as the
CGLMP inequality and also relevant to the CGLMP inequality.
%
%
For two-qutrit system, the CGLMP inequality reduces to:
\begin{widetext}
\begin{eqnarray}\label{CGLMP-Qrtrits}
I_{CGLMP}^{(3)}&=&[P(A_1\doteq B_1)+P(B_1\doteq A_2+1)+P(A_2\doteq
B_2)+P(B_2\doteq A_1)]
-[P(A_1\doteq B_1-1)\nonumber\\
&+&P(B_1\doteq A_2)+P(A_2\doteq B_2-1)+P(B_2\doteq A_1-1)]\}\leq2.
\end{eqnarray}
\end{widetext}
Here we denote the joint probability $P(A_i\doteq B_j+m)$ ($i, j=1,
2$) that the measurements $A_i$ and $B_j$ have outcomes that differ,
modulo three, by $m$: $P(A_i\doteq B_j+m)=\sum_{a=0}^{2}P(A_i=a,
B_j\doteq a+m)$. The inequality~(\ref{CGLMP-Qrtrits}) is
tight~\cite{L.Masanes} and it quantum violation is investigated in
Ref.~\cite{2006Chen-MaxV}.

To find a new Bell inequality, we can use the method introduced in
Sec. II. Let's denote the outcome of $A_i$ and $B_i$ by $a_i$ and
$b_i$ respectively. Unlike the qubits case, it is more convenient to
express the Bell function in terms of joint probability
$P(a_i+b_j\doteq r)$ that the measurements $A_i$ and $B_j$ have
outcomes that differ, modulo three, by $r$: $P(a_i+b_j\doteq
r)=\sum_{a=0,1,2}P(a_i=a,b_j\doteq r-a)$. 
Then the Bell function is defined as:
$\mathcal {B}(\lambda)=\sum_{\chi}c_{ijr}P(A_i+B_j\doteq r)$.
Here we specify $c_{ijr}=c_{jir}$ so that the Bell function is
symmetric under the permutations of $A_i$ and $B_i$. In fact, this
assumption is always valid since the two observers are in the same
status in the Bell-type scenario for derivation of Bell inequality.
Of course, the Bell inequalities corresponding to this kind of Bell
functions are symmetric under the permutations of $A_i$ and $B_i$,
too.  After a similar procedure as described in Sec. III, we find a
new three setting Bell inequality for two qutrits, which is relevant
to the CGLMP inequality for two qutrits (namely, the inequality
(\ref{CGLMP-Qrtrits})):
\begin{widetext}
\begin{eqnarray}\label{TSBIForQutrits-1}
I^{(3)}_{[3]}&=&-2P(a_1+b_1\doteq0)+P(a_1+b_1\doteq1)+P(a_1+b_1\doteq2)+P(a_1+b_2\doteq0)-P(a_1+b_2\doteq2)+P(a_2+b_1\doteq0)\nonumber\\
&-&P(a_2+b_1\doteq2)+P(a_1+b_3\doteq1)-P(a_1+b_3\doteq2)+P(a_3+b_1\doteq1)-P(a_3+b_1\doteq2)+P(a_2+b_3\doteq1)\nonumber\\
&-&P(a_2+b_3\doteq2)+P(a_3+b_2\doteq1)-P(a_3+b_2\doteq2)+P(a_3+b_3\doteq0)-P(a_3+b_3\doteq1)\leq4
\end{eqnarray}
\end{widetext}
It is easy to check that the inequality~(\ref{TSBIForQutrits-1}) is
tight and has four distinct \emph{``roots"}. To calculate the
quantum violation, we should note that the quantum prediction of the
joint probability $P(A_i=k, B_j=l)$ when $A_i$ and $B_j$ are
measured in the initial state $|\psi\rangle$ is given by
\begin{widetext}
\begin{eqnarray}\label{jointpro}
P(A_i=k,B_j=l)&=&|\langle kl|U(A)\otimes U(B)|\psi\rangle|^2\nonumber\\
&=&\text{Tr}\{[U(A)^{\dagger}\otimes
U(B)^{\dagger}]\hat{\Pi}_k\otimes\hat{\Pi}_l[U(A)\otimes
U(B)]|\psi\rangle\langle\psi|\}
\end{eqnarray}
\end{widetext}
where $U(A)$, $U(B)$ are the unitary transformation matrix and
$\hat{\Pi}_k=|k\rangle\langle k|$, $\hat{\Pi}_l=|l\rangle\langle l|$
are the projectors for systems A and B, respectively. Then for the
maximally entangled two-qutrit state:
$|\psi\rangle_3^{max}=\frac{1}{\sqrt{3}}(|00\rangle+|11\rangle+|22\rangle$,
the violation of inequality~(\ref{TSBIForQutrits-1}) is $5.1737$ and
its threshold visibility is $0.77314$, which is larger than that of
the CGLMP inequality for this state. This indicate that the
inequality (\ref{TSBIForQutrits-1}) is not as strong as the CGLMP
inequality. The maximal violation occurs at a nonmaximally entangled
state:
$|\phi\rangle_3^{nmax}=0.60297|00\rangle+0.5641|11\rangle+0.5641|22\rangle$
and the maximal violation is $5.1803$. However, numerical results
show that it is relevant to the CGLMP inequality. This is, to our
knowledge, the first tight three-setting Bell inequality for two
qutrits inequivalent to the CGlMP inequality. Using the same method,
we also find a three-setting Bell inequality for two qutrits, which
is as strong as the CGLMP inequality:
\begin{widetext}
\begin{eqnarray}\label{SameStrongToCGLMP}
I'^{(3)}_{[3]}&=&-P(a_1+b_1\doteq0)+P(a_1+b_1\doteq1)+4P(a_1+b_2\doteq0)-4P(a_1+b_2\doteq1)+4P(a_2+b_1\doteq0)\nonumber\\
&-&4P(a_2+b_1\doteq1)-3P(a_1+b_3\doteq1)+3P(a_1+b_3\doteq2)-3P(a_3+b_1\doteq1)+3P(a_3+b_1\doteq2)\nonumber\\
&+&7P(a_2+b_2\doteq0)-7P(a_2+b_2\doteq2)+3P(a_2+b_3\doteq1)-3P(a_2+b_3\doteq2)+3P(a_3+b_2\doteq1)\nonumber\\
&-&3P(a_3+b_2\doteq2)\leq14
\end{eqnarray}
\end{widetext}
For the maximally entangled state $|\psi\rangle_3^{max}$, the
quantum violation is $20.1105$ and the corresponding threshold
visibility is $0.6962$ which is the same as that of the CGLMP
inequality. Its maximal violation also occurs at a nonmaximally
entangled state:
$|\varphi\rangle_3^{nmax}=0.48876|00\rangle+0.61689|11\rangle+0.61689|22\rangle$
and the maximal violation is $20.1105$. Unfortunately, this
inequality is not tight. From our numerical results, there is no
three-setting BI for two qutrits that stronger than the CGLMP, just
like that there is no three-setting BI for two qubits which is
stronger than the CHSH inequality. A nature question arises: Can we
find a multi-setting Bell inequality for two qutrits that is more
stronger than the CGLMP inequality?

\section{Conclusion and remarks}
In summary, we generalized the method in ~\cite{2008Chen-HolderToBI}
to construct Bell inequality for qubits to higher dimensional
systems. Using the method, we have obtained some multi-setting tight
Bell inequalities, which are relevant to the CHSH inequality and the
CG inequality. Moreover, we present, for the first time, a three
setting tight Bell inequality for two qutrits, which is maximally
violated by nonmaximally entangled states and relevant to the CGLMP
inequality. The concept ``\emph{root}" seems very important for Bell
inequalities. It plays a vital role to construct and classify Bell
inequalities. Besides, from the numerical results, it seems like
that if two Bell functions have different numbers of distinct
``\emph{roots}", then the corresponding tight Bell inequalities may
be relevant to each other. A very important question related to the
generalized method is: can we use this method to find a general
tight Bell inequality for $N$-qudit system? We will investigate this
problem subsequently.

\end{document}